\documentclass[12pt]{article}

\usepackage[T2A]{fontenc}
\usepackage[utf8]{inputenc}
\usepackage[english]{babel}
\usepackage{graphicx}

\author{\bfseries Andrey Vasilyev}
\title{\bfseries Vector Potential and Magnetic  Field  of Axially Symmetric Currents}
\date{Retired from State Optical Institute, Saint Petersburg, Russia
e-mail <andrey@wavemech.org>}
\begin{document}

\maketitle

\begin{abstract}
\label{abstract}

A solution is proposed for finding the vector potential and magnetic field
of any distribution of currents with axial symmetry. In this approach, the
magnetic field and the vector potential are looked for not by solving a 
differential equation but rather through straightforward calculation of 
integrals of one scalar function. The solution is expressed in terms of 
the associated Legendre polynomials $P_{lm}$ with the index $m$ of the Legendre 
polynomials assuming one value only, $m=1$. The solution has the form of 
a series, with the coefficients of the polynomials being combinations of 
multipole moments.
 
\bigskip

{\bfseries Key words:} electrodynamics, vector potential, spherical harmonics,
Legendre polynomials, magnetic field.

\end{abstract}

\section{Introduction}

It is common knowledge that the scalar potential  $\Phi(\mathbf{r})$ of any 
charge distribution  $\rho(\mathbf{r'})$ can be calculated (using the Gaussian 
absolute system of units) by the relation (see, e.g., \cite{1}, Chapter 2)
\begin{equation}
\label{1}
\Phi(\mathbf{r})=\int\frac{\rho(\mathbf{r'})dV'}{|\mathbf{r}-\mathbf{r'}|}.
\end{equation}  

In the spherical coordinate system  $r, \vartheta, \varphi$ the  
$\Phi(\mathbf{r})$ potential 
can be conveniently found in many cases by expansion in terms of
spherical harmonics $Y_{lm}$ (see, e.g., \cite{2}). The potential at a point
is a sum of actions of all the charges involved. Therefore, if
we are interested in the potential at a point inside a charge 
distribution, we will have to split the region of integration
in Eq. (\ref{1}) into two parts by a sphere of radius $r$ centered at 
the pole $O$. As a result, we come to equalities of two types, 
for $r>r'$ and for  $r<r'$. Here  $r$ refers to the point with the
potential we are interested in, and  $r'$ is the point over which 
integration will be performed. Accordingly, the part of the 
potential formed by charges with coordinates $r'<r$ we shall 
denote with $\Phi_Q(\mathbf{r})$, and the other part, produced by charges
with coordinates  $r'>r$, with $\Phi_G(\mathbf{r})$. The expressions for 
$\Phi_Q(\mathbf{r})$
and $\Phi_G(\mathbf{r})$ written in the form appropriate for us here can be 
found, for instance, in  \cite{2} and are reproduced below:

\begin{equation}
\label{2}
\Phi_Q(\mathbf{r})=\sum_{l=0}^\infty\sum_{m=-l}^l\sqrt{\frac{4\pi}{2l+1}}\cdot
\frac{Q_{lm}(r)Y_{lm}(\vartheta, \varphi)}{r^{l+1}}  \qquad (r>r'),
\end{equation}
where $Q_{lm}(r)$ is a multipole moment of order $l, m$:

\begin{equation}
\label{3}
Q_{lm}(r)=\sqrt{\frac{4\pi}{2l+1}}\int_0^r\int\int\rho(\mathbf{r'})r'^lY_{lm}^*
(\vartheta',\varphi')dV',
\end{equation}

\begin{equation}
\label{4}
\Phi_G(\mathbf{r})=\sum_{l=0}^\infty\sum_{m=-l}^l\sqrt{\frac{4\pi}{2l+1}}
r^l{G_{lm}(r)Y_{lm}(\vartheta, \varphi)}  \qquad (r<r'),
\end{equation}
where $G_{lm}(r)$: 

\begin{equation}
\label{5}
G_{lm}(r)=\sqrt{\frac{4\pi}{2l+1}}\int_r^\infty\int\int\frac{\rho(\mathbf{r'})}
{r'^{l+1}}Y_{lm}^*(\vartheta',\varphi')dV'.
\end{equation}

The functions $Q_{lm}(r)$ and $G_{lm}(r)$ depend on $r$ on the upper and 
lower limits of integration as on a parameter.

The potential at a given point $\mathbf{r}$ is a sum of the potentials 
calculated for $r>r'$ and $r<r'$: 
$\Phi(\mathbf{r})=\Phi_Q(\mathbf{r})+\Phi_G(\mathbf{r})$.

The spherical functions $Y_{lm}$ can be written as (see, e.g., \cite{2}):

\begin{equation}
\label{6}
Y_{lm}(\vartheta,\varphi)=\delta_m\sqrt{\frac{2l+1}{4\pi}\cdot\frac{(l-|m|)!}
{(l+|m|)!}}P_{lm}(\cos\vartheta)e^{im\varphi}.
\end{equation}

In this expression, the integers  $l,m$ satisfy the conditions 
$l\ge 0$, $-l\le m\le l $; $\delta_m=(-1)^m$ for $m\ge 0$, $\delta_m=1$
for $m<0$.
$P_{lm}$ denotes here the associated Legendre polynomial. 

The vector potential $\mathbf{A}(\mathbf{r})$ of any current distribution 
$\mathbf{j}(\mathbf{r'})$ in vacuum can be calculated from the relation 
(see, e.g., \cite{1}, Chapter 2)

\begin{equation}
\label{7}
\mathbf{A}(\mathbf{r})=\frac{1}{c}\int\frac{\mathbf{j}(\mathbf{r'})dV'}
{|\mathbf{r}-\mathbf{r'}|},
\end{equation}
where $c$ is the electrodynamic constant.

Relation (\ref{7}) resembles in its form equality (\ref{1}), the difference 
being that integration is performed here not over the scalar but 
rather over the vector function $\mathbf{j}(\mathbf{r'})$. If one wanted
to employ the above method of potential calculation with the use of spherical
harmonics, one would have to apply all the calculations to each 
component separately, with subsequent vector addition of the results
thus obtained. This approach may meet with considerable difficulties.

On the other hand, if the currents have axial symmetry, one can suggest 
a fairly simple and straightforward method of calculation of the vector 
potential which reduces essentially to calculating an integral of one
scalar function only.

\section{Method of calculation of the vector potential of axially 
symmetric currents}

Assume a system of currents with a volume density $\mathbf{j}(\mathbf{r'})$
which have axial symmetry. Denote the axis of symmetry of 
the system with  $Z$.  Introduce a spherical coordinate system 
$r, \vartheta, \varphi$ such that the  $\vartheta=0$ axis coincides with 
the symmetry axis $Z$. Denote the unit vectors along the 
$r, \vartheta, \varphi$ axes by  $\mathbf{n}_r, \mathbf{n}_\vartheta, 
\mathbf{n}_\varphi$. 
The $\vartheta=\pi/2$ plane will be called the equatorial plane. 

The system being symmetric, the currents do not depend on the $\varphi$
coordinate and have only one component along it: 
 $\mathbf{j}(\mathbf{r'})=j(r',\vartheta')\mathbf{n}_{\varphi'}$.
This means that all $\mathbf{j}(\mathbf{r'})$ are circular currents, and 
that the vector of any element of the current lies in the plane parallel
to the equatorial plane (we shall call it the $S'$ plane). The symmetry 
of the current system suggests that the vector potential
$\mathbf{A}(\mathbf{r})$ is also axially symmetric, i.e., the magnitude of the 
vector potential $A(r,\vartheta)$ does not depend on coordinate $\varphi$. 
But this means that the magnitude of the vector potential $A(r,\vartheta)$ 
can be determined on any plane passing through the symmetry axis $Z$.
The simplest way appears to determine the potential on the plane 
passing through the $Z$ axis and the coordinate $\varphi=0$, and it is this
what we are going to do. Call this plane the $M$ plane.

Calculate the vector potential at a point $\mathbf{r}$ lying on the $M$ plane.
The coordinates of this point are  $(r,\vartheta,0)$. It is convenient for 
the purpose of this analysis to pass through this point a plane $S$
parallel to the equatorial plane. 

Consider our situation in more detail. Vector potential at a point 
$\mathbf{r}$ (lying on plane $M$) is produced by currents flowing throughout the 
space of interest here. Consider the potentials created by the current 
elements $d\mathbf{j}(\mathbf{r'})$. Figure 1 illustrates our case.

\begin{figure}
\includegraphics{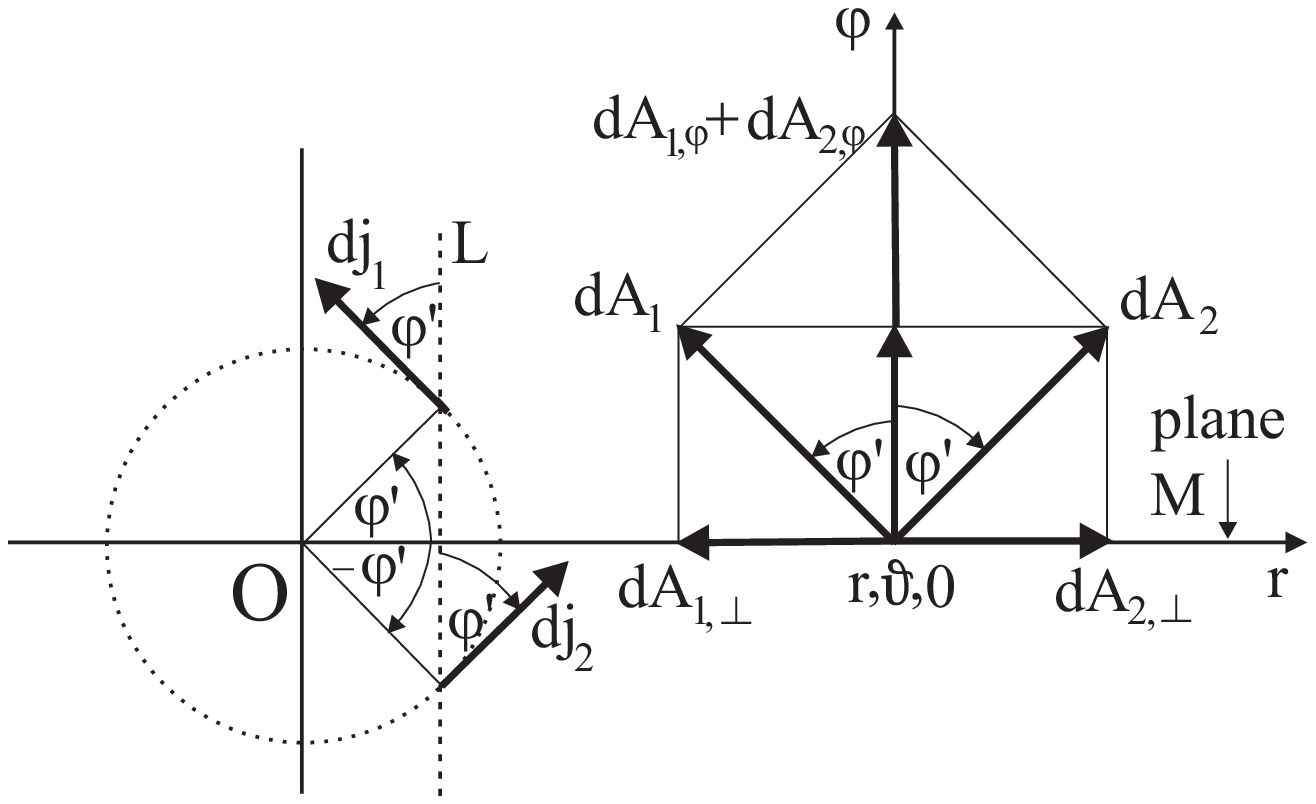}
\caption{}
\end{figure}

In Fig. 1, the origin of the spherical system of coordinates is at point $O$,
the $Z$ axis points directly at us, and the dashed line visualizes one of 
the circular current trajectories. This trajectory lies in the $S'$ plane.
The $S$ and $S'$ planes are parallel but do not coincide. The direction of
the $\varphi$ axis at point $(r,\vartheta,0)$ is specified in the figure. We are 
looking for the potential at this point. The $\varphi$ axis at point 
$(r,\vartheta,0)$ is perpendicular to the $M$ plane. 
The $r$ axis lies in the $M$ plane. 

The dashed line $L$ on the $S'$ plane is perpendicular to plane $M$,
i.e., parallel to the $\varphi$ axis at point $(r,\vartheta,0)$. 

Figure 1 shows two elements of the circular current with densities 
$d\mathbf{j}_1(\mathbf{r'}_1)$ and $d\mathbf{j}_2(\mathbf{r'}_2)$. 
The current element $d\mathbf{j}_1(\mathbf{r'}_1)$ produces at point 
$(r,\vartheta,0)$ on plane $M$ the potential $d\mathbf{A}_1(r,\vartheta,0)$.  
As seen from vector equality (\ref{7}), this potential is 
directed in space in the same way as the current element
$d\mathbf{j}_1(\mathbf{r'}_1)$. This means that the angles between a direction 
in space (for instance, straight lines perpendicular to plane $M$)
and vectors $d\mathbf{j}_1(\mathbf{r'}_1)$ and $d\mathbf{A}_1(\mathbf{r})$
should be equal (in Fig. 1, these are the $\varphi'$ angles).

The vector of the current element $d\mathbf{j}_1(\mathbf{r'}_1)$ is confined to
plane $S'$, which is parallel to the equatorial plane. As seen from
expression (\ref{7}), this means that the element of the vector potential
$d\mathbf{A}_1(\mathbf{r})$ generated by the current element 
$d\mathbf{j}_1(\mathbf{r'}_1)$ 
should likewise lie fully in the plane parallel to 
the equatorial plane, i.e., in plane $S$ (the $S$ and $S'$
planes do not coincide). Therefore, the $d\mathbf{A}_1(\mathbf{r})$ potential
can be resolved into two components, $dA_{1,\perp}$ and $dA_{1,\varphi}$
in the $S$ plane.  These components are shown in Fig. 1: 
 $dA_{1,\perp}=dA_1\sin\varphi'$ and $dA_{1,\varphi}=dA_1\cos\varphi'$,
 where the angle $\varphi'$
is the coordinate of the current element $d\mathbf{j}_1(\mathbf{r'}_1)$.

Since, however, the current is axially symmetric, there will always be 
an element of current $d\mathbf{j}_2$ equal in magnitude to the element 
$d\mathbf{j}_1$ and located at the same distance from point $(r,\vartheta,0)$
as the  $d\mathbf{j}_1$ element. The coordinates of this current element are 
$(r',\vartheta',-\varphi')$. It produces at the same point  $(r,\vartheta,0)$
the potential $d\mathbf{A}_2(r,\vartheta,0)$. Resolving potential 
$d\mathbf{A}_2(r,\vartheta,0)$ in the  $S$ plane into two components, 
$dA_{2,\perp}$ and $dA_{2,\varphi}$, we see immediately that the components 
$dA_{1,\perp}$ and  $dA_{2,\perp}$ cancel, while $dA_{1,\varphi}$ and
$dA_{2,\varphi}$ add (see Fig. 1). Thus, after integration over all elements
of current $d\mathbf{j}(\mathbf{r'})$ the potential $\mathbf{A}(\mathbf{r})$
will have only one component, the one along the $\varphi$ axis, left. 
But this implies that in determination of the $\mathbf{A}(\mathbf{r})$
potential one may, rather than taking into account the whole potential 
$d\mathbf{A}(\mathbf{r})$ produced by the current element 
$d\mathbf{j}(\mathbf{r'})$, restrict oneself to the projection 
$dA_\varphi$ of this potential onto the $\varphi$ axis: 
$dA_\varphi=dA\cos\varphi'$, where $\varphi'$ is the coordinate of the
element of current $d\mathbf{j}(\mathbf{r'})$. Said otherwise, the 
potential $dA_\varphi$ is generated not by the current element
$d\mathbf{j}(\mathbf{r'})$ as a whole but by the projection of this 
element on the line $L$ parallel to vector $\varphi$ at point $(r,\vartheta,0)$.

We can now, on replacing vector $\mathbf{j}(\mathbf{r'})$ in Eq. (\ref{7})
with a scalar 
$J(\mathbf{r'})=J(r',\vartheta',\varphi')=j(r',\vartheta')cos\varphi'$,
 calculate the magnitude of the vector potential

\begin{equation}
\label{8}
A(r,\vartheta,0)=\frac{1}{c}\int\frac{J(\mathbf{r'})dV'}
{|\mathbf{r}-\mathbf{r'}|}=\frac{1}{c}\int\frac{j(r',\vartheta')\cos\varphi'dV'}
{|\mathbf{r}-\mathbf{r'}|}.
\end{equation}
The potential at any point $\mathbf{r}$ can now be written as

\begin{equation}
\label{9}
\mathbf{A}(\mathbf{r})=A(r,\vartheta,0)\mathbf{n}_\varphi.
\end{equation}

Because expression (\ref{8}) contains scalar quantities, in calculation 
of the integral one may resort to expanding the integrand in spherical
harmonics. Note, however, that because expression (\ref{8}) contains 
$\cos\varphi'$ as a factor, one may conveniently use in expressions 
for $Y_{lm}$ in place of exponential functions employed customarily, 
trigonometric ones (see. e.g., \cite{3}, Chapter 21). In this case 
one will have to replace expression (\ref{6}) with the following relations:

\begin{equation}
\label{10}
Y_{lm}^c(\vartheta,\varphi)=\sqrt{\frac{2l+1}{2\pi}\cdot\frac{(l-m)!}
{(l+m)!}}P_{lm}(\cos\vartheta)\cos{m\varphi},
\end{equation}

\begin{equation}
\label{11}
Y_{lm}^s(\vartheta,\varphi)=\sqrt{\frac{2l+1}{2\pi}\cdot\frac{(l-m)!}
{(l+m)!}}P_{lm}(\cos\vartheta)\sin{m\varphi},
\end{equation}
$$
(l=0, 1, 2, ...; m=0, 1, 2, ..., l).
$$
This involves a minor change in the form of expansion of the function.

\begin{flushleft}
For $r>r'$:
\end{flushleft}

\begin{equation}
\label{12}
A_Q(r,\vartheta,0)=\frac{1}{c}\sum_{l=0}^\infty\sqrt{\frac{4\pi}{2l+1}}\left
(\frac{1}{2}
\frac{Q_{l0}^cY_{l0}^c}{r^{l+1}}+\sum_{m=1}^l\frac{Q_{lm}^cY_{lm}^c}{r^{l+1}}+
\sum_{m=1}^l\frac{Q_{lm}^sY_{lm}^s}{r^{l+1}}\right),
\end{equation}
where the multipole moments $Q_{lm}^c$ and $Q_{lm}^s$ have the form

\begin{equation}
\label{13}
Q_{lm}^c(r)=\sqrt{\frac{4\pi}{2l+1}}\int_0^r\int\int J(\mathbf{r'})r'^lY_{lm}^c
(\vartheta',\varphi')dV',
\end{equation}

\begin{equation}
\label{14}
Q_{lm}^s(r)=\sqrt{\frac{4\pi}{2l+1}}\int_0^r\int\int J(\mathbf{r'})r'^lY_{lm}^s
(\vartheta',\varphi')dV'.
\end{equation}
For $r<r'$:

\begin{equation}
\label{15}
A_G(r,\vartheta,0)=\frac{1}{c}\sum_{l=0}^\infty\sqrt{\frac{4\pi}{2l+1}}
\left(\frac{1}{2}r^l
G_{l0}^cY_{l0}^c+\sum_{m=1}^lr^lG_{lm}^cY_{lm}^c+
\sum_{m=1}^lr^lG_{lm}^sY_{lm}^s\right),
\end{equation}
with the multipole moments $G_{lm}^c$ and $G_{lm}^s$

\begin{equation}
\label{16}
G_{lm}^c(r)=\sqrt{\frac{4\pi}{2l+1}}\int_r^\infty\int\int\frac
{J(\mathbf{r'})}{r'^{l+1}}Y_{lm}^c(\vartheta',\varphi')dV',
\end{equation}

\begin{equation}
\label{17}
G_{lm}^s(r)=\sqrt{\frac{4\pi}{2l+1}}\int_r^\infty\int\int\frac
{J(\mathbf{r'})}{r'^{l+1}}Y_{lm}^s(\vartheta',\varphi')dV'.
\end{equation}

The function $\cos\varphi'$ entering the expression for $J(\mathbf{r'})$
also belongs to the system of functions $Y_{lm}^c$ (see expression (\ref{10})).
Therefore, because of the functions $\cos{m\varphi'}$ and $\sin{m\varphi'}$ 
being orthogonal, part of the multipole moments will vanish in the course 
of integration of expressions (\ref{13}), (\ref{14}) and (\ref{16}), (\ref{17})
(orthogonality in index $m$). The only terms left will be those with $m=1$,
i.e., the terms $Q_{l1}^c(r)$ and $G_{l1}^c(r)$. As a result, the double 
series (\ref{12}) and (\ref{15}) will become single series and look as

\begin{equation}
\label{18}
A_Q(r,\vartheta,0)=\frac{1}{c}\sum_{l=1}^\infty\sqrt{\frac{4\pi}{2l+1}}
\frac{Q_{l1}^cY_{l1}^c}{r^{l+1}}
\qquad \mbox{for } r>r', 
\end{equation}

\begin{equation}
\label{19}
A_G(r,\vartheta,0)=\frac{1}{c}\sum_{l=1}^\infty\sqrt{\frac{4\pi}{2l+1}}
G_{l1}^cY_{l1}^cr^l
\qquad \mbox{for } r<r'. 
\end{equation}

The next step appears to be substitution in these series of the expression
for $Y_{l1}^c$ from formula (\ref{10}). We have to remember, however, 
that we are looking for the potential on the $\varphi=0$ plane only.
For all other values of the $\varphi$ coordinate, the relations we have
derived will yield wrong values of the potential. Therefore, prior to 
substituting the expression for $Y_{l1}^c$ from relation (\ref{10}), 
we will have to zero the coordinate $\varphi$. We obtain naturally
$\cos\varphi=1$. Now the potentials take on the form

\begin{equation}
\label{20}
A_Q(r,\vartheta,0)=\frac{1}{c}\sum_{l=1}^\infty\frac{2\pi}{l(l+1)}
\frac{Q^A_{l1}(r)P_{l1}(\cos\vartheta)}{r^{l+1}}
\qquad \mbox{for } r>r',
\end{equation}
where the multipole moments $Q^A_{l1}$ are

\begin{equation}
\label{21}
Q^A_{l1}(r)=
\int_0^r\int_0^\pi j(r',\vartheta')r'^lP_{l1}(\cos\vartheta')r'^2\sin\vartheta' 
d\vartheta' dr',
\end{equation}

\begin{equation}
\label{22}
A_G(r,\vartheta,0)=\frac{1}{c}\sum_{l=1}^\infty\frac{2\pi}{l(l+1)}
r^lG^A_{l1}(r)P_{l1}(\cos\vartheta)
\qquad \mbox{for } r<r',
\end{equation}
with the multipole moments $G^A_{l1}$

\begin{equation}
\label{23}
G^A_{l1}(r)=
\int_r^\infty\int_0^\pi \frac{j(r',\vartheta')}{r'^{l+1}}P_{l1}(\cos\vartheta')
r'^2\sin\vartheta' d\vartheta' dr'.
\end{equation}

In these formulas, $j(r',\vartheta')$ is the circular current density,
and $Q^A_{l1}(r)$ and $G^A_{l1}(r)$  are the parts of the multipole moments 
left after integration over the $\varphi$ coordinate (truncated multipole 
moments). The vector potential at any point in space can be written as

\begin{equation}
\label{24}
\mathbf{A}(\mathbf{r})=[A_Q(r,\vartheta,0)+A_G(r,\vartheta,0)]
\mathbf{n}_\varphi,
\end{equation}
whence we finally come to
\medskip

\fbox{
\parbox{12.8cm}{
\begin{equation}
\label{25}
\mathbf{A}(\mathbf{r})=\frac{1}{c}\sum_{l=1}^\infty\frac{2\pi}{l(l+1)}
\left(\frac{Q^A_{l1}(r)}{r^{l+1}}+r^lG^A_{l1}(r)\right)
P_{l1}(\cos\vartheta)\mathbf{n}_\varphi.
\end{equation}
}}
\medskip

This is the final expression for the vector potential $\mathbf{A}(\mathbf{r})$.
As evident from Eq. (\ref{25}), vector potential $\mathbf{A}(\mathbf{r})$
has only one component, $A_\varphi$.

Now we impose an additional constraint, namely, that the currents are 
symmetrical with respect to the equatorial plane (which quite often is 
the case). In this case, expression (\ref{25}) may contain only functions 
$P_{l1}$ symmetric about the equator, i.e., $P_{2n+1,1}$ functions. 
Now expression (\ref{25}) for the vector potential has to be replaced by

\begin{equation}
\label{26}
\mathbf{A}(\mathbf{r})=\frac{1}{c}\sum_{n=0}^\infty\frac{\pi}{(n+1)(2n+1)}
\left(\frac{Q^A_{2n+1,1}(r)}{r^{(2n+1)+1}}+r^{2n+1}G^A_{2n+1,1}(r)\right)
P_{2n+1,1}(\cos\vartheta)\mathbf{n}_\varphi.
\end{equation}

It may be of interest to consider a particular relevant case. Consider
axially symmetric currents with a  $\sin\vartheta$ dependence on the 
$\vartheta$ angle. This case becomes realized, for instance, if the currents 
are generated by rotation of charges, with all the charges in a layer 
of thickness $dr$ distributed uniformly over the angle  $\vartheta$ 
and rotating with the same angular velocity.

The $\sin\vartheta$ function belongs to the system of associated Legendre 
functions. Therefore, the $P_{l1}$ functions being orthogonal with respect 
to index $l$, the larger part of the terms in the $Q^A_{l1}(r)$ and
$G^A_{l1}(r)$ expansions (see expressions (\ref{21}) and (\ref{23})) 
will vanish in calculation, with only $Q^A_{11}(r)$ and $G^A_{11}(r)$
terms with index $l=1$ retained. In this case, the potential will come 
out as a sum of two terms only:

\begin{equation}
\label{27}
\mathbf{A}(\mathbf{r})=\frac{\pi}{c}\left
(\frac{Q^A_{11}(r)}{r^2}+rG^A_{11}(r)\right)\sin\vartheta\mathbf{n}_\varphi.
\end{equation} 

\section{Magnetic fields of axially symmetric currents}

Magnetic field $\mathbf{H}(\mathbf{r})$ can be calculated with the general
expression $\mathbf{H}(\mathbf{r})=\nabla\times\mathbf{A}(\mathbf{r})$. 
The vector potential of axially symmetric currents has only one component,
$A_\varphi$:
$\mathbf{A}(\mathbf{r})=A(r,\vartheta)\mathbf{n}_\varphi$.
With this in mind, and using the general expression for
$\nabla\times\mathbf{A}(\mathbf{r})$ in the spherical coordinate system 
(see, for instance, \cite{3}, Chapter 6)

$$
\nabla\times\mathbf{A}=\left[\frac{1}{r\sin\vartheta} \left(\frac{\partial}
{\partial\vartheta}(\sin\vartheta A_\varphi)-\frac{\partial A_\vartheta}
{\partial\varphi}\right)\right]\mathbf{n}_r+
$$
$$
\left[\frac{1}{r\sin\vartheta}\frac{\partial A_r}{\partial\varphi}-
\frac{1}{r}\left(\frac{\partial}{\partial r}(rA_\varphi)\right)\right]
\mathbf{n}_\vartheta+
\left[\frac{1}{r}\frac{\partial(rA_\vartheta)}{\partial r}-\frac{1}{r}
\frac{\partial A_r}{\partial\vartheta}\right]\mathbf{n}_\varphi,
$$
we finally come to 

\begin{equation}
\label{28}
\mathbf{H}=\nabla\times\mathbf{A}=\left(\frac{1}{r}\frac{\cos\vartheta}
{\sin\vartheta}A_\varphi+\frac{1}{r}\frac{\partial}{\partial\vartheta}
A_\varphi\right)\mathbf{n}_r-
\left(\frac{1}{r}A_\varphi+\frac{\partial}{\partial r}A_\varphi\right)
\mathbf{n}_\vartheta.
\end{equation} 
with  $A_\varphi$ expressed by formula (\ref{25}). 

Consider the expression $\frac{\partial}{\partial r}A_\varphi$ in any of 
the terms of expansion (\ref{25}):

\begin{equation}
\label{29}
\frac{\partial}{\partial r}A_\varphi=\frac{1}{c}\frac{2\pi}{l(l+1)}
\left(\frac{-(l+1)}{r^{l+2}}Q^A_{l1}+lr^{l-1}G^A_{l1}+
\frac{1}{r^{l+1}}\frac{\partial}{\partial r}Q^A_{l1}+r^l\frac{\partial}
{\partial r}G^A_{l1}\right)P_{l1}(\cos\vartheta).
\end{equation} 
By the theorem of Leibniz--Newton (see, e.g., \cite{3}, Chapter 4, item 4.6-5):

$$
\frac{d}{dx}\int_a^xf(t)dt=f(x),
$$
the last two terms in expression (\ref{29}) cancel, with the final
expression taking on the form
\newpage

$$
\mathbf{H}(\mathbf{r})=\frac{1}{c}\sum_{l=1}^\infty\frac{2\pi}{l(l+1)}
\left\{\left[\left(\frac{Q^A_{l1}(r)}{r^{l+2}}+r^{l-1}G^A_{l1}(r)\right)
\right.\right.\cdot
$$

\begin{equation}
\label{30}
\left.\left(\frac{\cos\vartheta}{\sin\vartheta}P_{l1}(\cos\vartheta)+
\frac{\partial P_{l1}(cos\vartheta)}
{\partial\vartheta}\right)\right]\mathbf{n}_r+
\end{equation}
$$
\left.\left[\left(\frac{lQ^A_{l1}(r)}{r^{l+2}}-(l+1)r^{l-1}G^A_{l1}(r)\right)
P_{l1}(\cos\vartheta)\right]\mathbf{n}_\vartheta\right\}.
$$

If the currents are symmetric about the equator, expression (\ref{30}), 
similar to formula (\ref{26}), will contain only equatorially symmetric 
functions $P_{2n+1,1}$, and summation over $n$ should be performed from 
$0$ to $\infty$.

Consider a few particular cases. In the case covered by potential (\ref{27}),
the magnetic field should be written as

\begin{equation}
\label{31}
\mathbf{H}=\frac{\pi}{c}\left[\left(\frac{Q^A_{11}}{r^3}+
G^A_{11}\right)2\cos\vartheta\mathbf{n}_r+
\left(\frac{Q^A_{11}}{r^3}-2G^A_{11}\right)\sin\vartheta\mathbf{n}_
\vartheta\right].
\end{equation} 

To illustrate the use of this expression, consider the following simple 
problem: rotation of a charged sphere about one of the diameters. Let the 
sphere of radius $R$, with charge  $q$ distributed uniformly over its surface,
rotate about the $Z$ axis with an angular velocity  $\mathbf{w}$. Calculate 
the magnetic field and the vector potential inside and outside the sphere.

The density of the surface current produced by rotation of the sphere can 
be expressed in the following way:

\begin{equation}
\label{32}
\mathbf{j}=\frac{qw}{4\pi R}\delta(r-R)\sin\vartheta\mathbf{n}_\varphi.
\end{equation} 

Substituting the expression for current density (\ref{32}) in relations 
(\ref{16}) and (\ref{17}), we come to the following expressions for the 
(truncated) multipole moments

$$
 Q^A_{11}=0, \qquad G^A_{11}=\frac{qw}{3\pi R} \qquad \mbox{for } r<R,
$$ 

$$
 Q^A_{11}=\frac{qwR^2}{3\pi}, \qquad G^A_{11}=0 
 \qquad \mbox{for } r>R.
$$

Now the magnetic field and the vector potential acquire their final form

\begin{equation}
\label{33}
\mathbf{H}=\frac{2qw}{3cR}(\cos\vartheta\mathbf{n}_r-\sin\vartheta\mathbf{n}
_\vartheta), \qquad  
\mathbf{A}=\frac{qwr}{3cR}\sin\vartheta\mathbf{n}_\varphi
\qquad \mbox{for } r<R,
\end{equation} 

\begin{equation}
\label{34}
\mathbf{H}=\frac{2qwR^2}{3cr^3}(\cos\vartheta\mathbf{n}_r+\frac{1}{2}
\sin\vartheta\mathbf{n}_\vartheta), \qquad
\mathbf{A}=\frac{qwR^2}{3cr^2}\sin\vartheta\mathbf{n}_\varphi
\qquad \mbox{for } r>R.
\end{equation}

Significantly, this technique permits one to determine the magnetic field
without prior calculation of the vector potential.

The solution to this problem can be found, for instance, in monographs 
\cite{1} (problem 2.85) and \cite{2} (problem 253):

$$
\mathbf{H}=\frac{2q\mathbf{w}}{3cR} \qquad \mbox{for }r<R, \qquad
\mathbf{H}=\frac{3\mathbf{r}(\mathbf{m}\cdot\mathbf{r})}{r^5}-\frac{\mathbf{m}}
{r^3} \qquad \mbox{for }r>R,
$$
where $\mathbf{m}=\frac{qR^2}{3c}\mathbf{w}$ is the magnetic moment of 
the system. One can readily verify that these expressions are identical. 
In monographs  \cite{1} and \cite{2} it is proposed to solve this problem
by differential calculus. Our approach, by contrast, has yielded the answers 
to it by straightforward calculation.

\end{document}